 \documentclass[12pt,preprint]{aastex}

\slugcomment{Not to appear in Nonlearned J., 45.}

\shorttitle{A Very Close Binary Black Hole}
\shortauthors{Iguchi, Okuda, Sudou}

\begin{document}


\title{A Very Close Binary Black Hole in a Giant Elliptical Galaxy 3C 66B \\
     and its Black Hole Merger}


\author{Satoru Iguchi\altaffilmark{1}, 
Takeshi Okuda\altaffilmark{2}, 
and 
Hiroshi Sudou\altaffilmark{3}}


\altaffiltext{1}{National Astronomical Observatory of Japan, 2-2-1 Osawa, Mitaka, Tokyo 181-8588, Japan; s.iguchi@nao.ac.jp}
\altaffiltext{2}{Nagoya University, Furo-cho, Chikusa-ku, Nagoya, Aichi 464-8602, Japan; okuda@a.phys.nagoya-u.ac.jp}
\altaffiltext{3}{Gifu University, 1-1 Yanagido, Gifu, Gifu 501-1193, Japan; sudou@gifu-u.ac.jp}


\begin{abstract}
Recent observational results 
provide possible evidence 
that
binary black holes (BBHs) exist in the center of giant galaxies 
and may merge to form a supermassive black hole in the process of their evolution. 
We first detected a periodic flux variation on a cycle of $93\pm1$ days 
from the 3-mm monitor observations of a giant elliptical galaxy \object{3C 66B} 
for which an orbital motion with a period of $1.05\pm0.03$ years had been already observed. 
The detected signal period being shorter than the orbital period can be explained 
by taking into consideration the Doppler-shifted modulation due to the orbital motion of a BBH. 
Assuming that the BBH has a circular orbit 
and 
that the jet axis is parallel to the binary angular momentum,
our observational results demonstrate the presence of a very close BBH 
that has the binary orbit with 
an orbital period of $1.05\pm0.03$ years, 
an orbital radius of $(3.9\pm1.0) \times 10^{-3}$ pc, 
an orbital separation of $(6.1^{+1.0}_{-0.9}) \times 10^{-3}$ pc,
the larger black hole mass of $(1.2^{+0.5}_{-0.2}) \times 10^9$ $M_{\sun}$, 
and the smaller black hole mass of $(7.0^{+4.7}_{-6.4}) \times 10^8$ $M_{\sun}$. 
The BBH decay time of $(5.1^{+60.5}_{-2.5})\times 10^2$ years 
provides evidence for the occurrence of black hole mergers. 
This Letter will demonstrate the interesting possibility of black hole collisions 
to form a supermassive black hole in the process of evolution, 
one of the most spectacular natural phenomena in the universe.
\end{abstract}


\keywords{black hole physics --- galaxies: active --- galaxies: elliptical and lenticular, cD --- galaxies: formation --- galaxies: individual(3C 66B)}



\section{Introduction}

IThe major galaxy mergers in galaxy formation inevitably 
leads to the presence of a binary black hole (BBH) 
located in the center of a giant galaxy \citep{begelman80,yu02}. 
The following observational results provide possible evidence of 
the presence of BBHs: double compact cores \citep{rodriguez06}, 
two broad-line emission systems \citep{boroson09}, 
the X-shaped morphology of radio lobes \citep{merritt02}, 
wiggle patterns of radio jets \citep{roos93,abraham98,lobanov05}, 
periodic variations and periodic outbursts \citep{sillanpaa88,valtaoja00,valtonen08}, 
and the orbital motion of the compact core \citep{sudou03}. 
Strong evidence for BBHs as linking some observational results, 
however, has not yet been found. 
Clarifying the mechanism for the formation of BBHs is essential 
in the study of galaxy mergers in galaxy formation \citep{sanders88,treister10}, 
in the role of black hole mergers in the evolution of supermassive black holes \citep{baker06}, 
and in the detection of gravitational waves at the phase of BBH orbital decay \citep{begelman80,yu02,baker06}.

\object{3C 66B} is a nearby low-luminosity FR I radio galaxy and 
giant elliptical galaxy with a redshift ($z$) of 0.0213 \citep{huchra99}, 
and has the strong jet and counter jet that extend to 
about 100 kiloparsecs (kpc) which have been observed 
at radio \citep{hardcastle96}, infrared \citep{tansley00}, 
optical \citep{macchetto91} and X-ray waves \citep{hardcastle01}. 
The orbital motion of the compact core in \object{3C 66B}, 
which has a period of $1.05\pm0.03$ years, was observed 
with a position accuracy of 10 micro arcseconds ($\mu$as) 
by using phase-referencing Very Long Baseline Interferometry (VLBI) \citep{sudou03}. 

Several observational and theoretical studies have shown 
that the periodic flux variation and periodic outburst activities 
in the radio, optical, X-ray, and  $\gamma$-ray light curves 
are associated with the orbital motion of a BBH 
in the center of the galaxies \citep{rieger00,depaoils02,depaoils04}. 
Here we report the detection of a signal periodicity 
in light curves from the compact core of \object{3C 66B}, 
which indicates the presence of a very close BBH in the center of this object,
and also present evidence for black hole mergers. 
We use a Hubble constant ($H_0$) of 71 km s$^{-1}$ Mpc$^{-1}$, 
the matter density ($\Omega_\mathrm{M}$) of 0.27 
and the vacuum energy ($\Omega_\mathrm{\Lambda}$) of 0.73 
in this Letter, 
resulting that an angular size or separation of 1 milliarcsecond (mas) corresponds to 0.436 pc 
at the distance of 3C 66B.

\section{Observations, Data Reduction, and Results}

The millimeter-wavelength flux variation for \object{3C 66B} 
was monitored every two weeks from the end of 2003 through to 
the beginning of 2005 at 93.716 GHz using the Nobeyama Millimeter Array (NMA) of 
the National Astronomical Observatory of Japan (NAOJ), 
and every four weeks from the end of 2005 through to the middle of 2006 at 86.2 GHz
using the Plateau de Bure Interferometer (PdBI) of 
the Institut de Radioastronomie Millim\'{e}trique (IRAM). 
These flux monitor observations using two-type millimeter arrays 
will enable us to eliminate any dependence on the characteristic features of each array, 
giving us more reliable and accurate results.

In the NMA observation, \object{3C 84} and \object{0133+476} 
were observed as a visibility calibrator. 
Also, the flux densities of each calibrator were precisely derived by observing \object{Uranus} and \object{Neptune}. 
Phase fluctuations in the observation are caused by short-term variations 
of the excess path length in the atmosphere, which is mainly due to the time variation of water vapor pressure. 
Since these phase variations cause decoherence in the visibility data, 
it is necessary to correct this loss. 
The decoherence factors at all observation epochs were estimated 
from the synthesized maps of each visibility calibrator \citep{kohno05}. 
To accurately derive the flux densities of the observed object, 
the flux data at epochs with a decoherence of more than 20 \% were flagged. 
\object{3C 454.3} and \object{0420$-$014} were observed for the bandpass calibration. 
The weighted mean of the flux density data that were estimated 
from each visibility calibrator was plotted. 
In the PdBI observation, the flux densities of \object{3C 66B} were corrected 
by observing \object{3C 84} and the compact H$_\mathrm{II}$ region, \object{MWC 349} and/or \object{CRL 618}, 
except \object{3C 84} on January 14 and \object{MCW 349} on July 23 in 2006. 
The phase and bandpass calibrations were performed by \object{3C 66B} itself.

We first investigated the contribution of large-scale jet emissions 
into observed flux densities. 
A map of \object{3C 66B} with millimeter jets can be made 
by combining the visibility data obtained from the NMA monitoring observations 
for all epochs (see Figure 1a). 
The total flux density of the millimeter map of 3C 66B at 93 GHz exceeds 500 mJy, 
while the sum of the expanding jet and counter jet is less than 24 mJy. 
We made the Spectral Energy Distributions (SEDs) of the jet and the counter jet 
from our data and other data at different wavelengths, respectively. 
We found that these SEDs follow (agree with) a simple power-law synchrotron model, 
even though the flux densities at all wavelengths were observed 
at the different dates (see Figure 1b). 
This fact indicates that the observed flux variations of \object{3C 66B} were 
dominated by the unresolved core, 
not by the expanding jet and counter jet. 

Figure 2 shows the 3-mm peak flux monitor data obtained from 
the unresolved core of \object{3C 66B} between 2003 November and 2006 August, 
and the spectral analysis results from these monitor data 
by using Lomb-Scargle periodogram \citep{lomb76,scargle82}. 
From these results, we detected a periodic flux variation on a cycle of 93$\pm$1 days 
with a maximum-to-minimum amplitude ratio of $1.19\pm0.03$, 
and non-periodic flux variation from the unresolved core. 
The variability timescale of the observed signal is shorter than 
that of the orbital period of about 1 year. 
Due to the orbital motion around the center of gravity in a BBH, 
the Doppler factor for the emitting region is 
a periodical function of time \citep{rieger00,depaoils02,depaoils04}, and 
the observed signal periodicity has a geometrical origin due to Doppler-shifted modulation. 
Compared to the intrinsic period, the observed signal period is shortened 
by the relativistic effect \citep{camenzind92}. 
It is also noted that the non-periodic flux variation may be 
caused by the non-thermal radiation in the flaring state that is emitted 
by a relativistic emission region not directly affected by a BBH 
(e.g., a shock in the knots of jet in the core that is not resolved 
with the angular resolutions of NMA and PdBI).

\section{Discussion and Summary}

Both the observed flux variation (see Figure 2) and the orbital motion \citep{sudou03} were well 
fitted by a simple model of sin function. 
This result indicates that the difference between the apogee and perigee velocities will be small, 
and that the Doppler boosting factor will be small. 
If they were large, 
the profile of the observed flux variation would be that of a spike 
like the periodic outburst activities in the flaring state. 
This indicates that the binary system in \object{3C 66B} has a nearly circular orbit. 
As a consequence of the binary orbital revolution around the center of gravity, 
the observed signal period ($P_\mathrm{obs}$) is related to 
the orbital period ($P_\mathrm{k}$) by \citep{rieger00}
\begin{equation}
P_\mathrm{obs} = (1 + z)(1 - \beta \cos i) P_\mathrm{k},
\label{eq:period}
\end{equation}
where $\beta$ is $v/c$ with the jet outflow velocity of $v$ 
and the light speed of $c$, 
and $i$ is the inclination angle between the jet axis and the line of sight. 
The apparent speed in the relativistic jet is written by \citep{ginzburg96} 
\begin{equation}
\beta_\mathrm{app} = (\beta \sin i) / (1 - \beta \cos i). 
\label{eq:apparent}
\end{equation}
Given the detected signal period ($P_\mathrm{obs}$) of 93$\pm$1 days, 
the orbital period ($P_\mathrm{k}$) of $1.05\pm0.03$ years \citep{sudou03}, and 
the derived apparent speed ($\beta_\mathrm{app}$) of 
0.30$\pm$0.04 (see Figure 3), 
the following physical parameters for an inner-jet outflow 
that has the effect of the BBH orbital motion found in \object{3C 66B} were estimated 
from equations (\ref{eq:period}) and (\ref{eq:apparent}); 
a speed ($\beta$=$v/c$) of 0.77$\pm$0.01 
and an inclination angle ($i$) of 5.3$\pm$0.8 degrees 
in accordance with the standard model for jet bulk motion. 
Consequently, the derived Doppler factor ($\delta$) of 2.8 
[=$(1- \beta^2)^{1/2}/(1- \beta \cos i)$] is a small value that does not 
have a strong effect on the profile of the periodical flux variation.

The jet is presumably linked to the accretion disk around a black hole. 
At least one of the two black holes in a binary system will have an accretion disk. 
If both of the black holes have an accretion disk, 
the relationship between the observed jet and the two black holes is not simple. 
Assuming that the observed jet is dominated by the accretion disk 
around either one of the two black holes, we 
show two schematic geometries of a BBH in Figure 4. 
The orbital motion velocity ($v_\mathrm{bl}$) around the center of gravity 
can be written as $\omega R \sin(\omega t)$, 
where $\omega=2\pi / P_\mathrm{k}$ is the intrinsic orbital frequency. 
The observed flux modulation by Doppler boosting can be written by 
\begin{equation}
S(\nu) = \delta(t)^{3+\alpha} S^*(\nu), 
\label{eq:flux}
\end{equation}
where $S^*(\nu)$ is the intrinsic spectral flux density in the comoving frame, 
$\alpha$ is the source spectral index ($S \propto \nu^{-\alpha}$), 
and $\delta(t)$ is the Doppler factor including the time variation 
due to the orbital motion of the BBH, which is written by \citep{depaoils02,depaoils04} 
\begin{equation}
\delta (t)=\frac{\sqrt{1 - (\beta^2 + \beta_\mathrm{bl}(t)^2)}}{1- (\beta \cos i + \beta_\mathrm{bl}(t) \sin i)},
\label{eq:doppler}
\end{equation}
where $\beta_\mathrm{bl}(t)$ is $v_\mathrm{bl}(t)/c$. 
From equation (\ref{eq:flux}), the observed maximum-to-minimum flux ratio 
can be expressed as 
\begin{equation}
\left( \frac{\delta_\mathrm{max}}{\delta_\mathrm{min}}\right)^{3+\alpha} = \frac{S_\mathrm{max}}{S_\mathrm{min}}. 
\label{eq:ratio}
\end{equation}
By applying this model to the sin curve of the periodic flux variation 
caused by a BBH (see Figure 2), 
an orbital motion velocity ($v_\mathrm{bl}$) of the BBH can be derived 
from equations (\ref{eq:doppler}) and (\ref{eq:ratio}), 
and an orbital radius ($R$) of $(1.2\pm0.3)\times10^{16}$ cm 
[$(3.9\pm1.0)\times10^{-3}$ pc] can then be estimated when $\alpha$=0. 
Although an upper limit for the orbital radius could only be estimated 
from the core motion using phase-referencing VLBI \citep{sudou03}, 
the orbital radius of a BBH can be directly derived 
by adding the periodic flux monitor data that we found successfully.

From two types of geometry of a BBH system shown in Figure 4, 
an orbital separation ($d$) of two black holes is $R (1+q)$ (see Figure 4a) or $R (1+q)/q$ (see Figure 4b) 
when $q = m/M$, where $m$ and $M$ denote the mass of the smaller and larger black holes, respectively. 
Since the revolution period of a two-body problem with a circular orbit 
has a relational expression of $P_\mathrm{k} = 2\pi G^{-1/2} d^{3/2} (M+m)^{-1/2}$ 
(where $G$ is the gravitational constant), 
the mass of the larger black hole ($M$) is estimated to be 
\begin{equation}
(2\pi)^2P_\mathrm{k}^{-2} G^{-1}R^3 (1+q)^2
\label{eq:small}
\end{equation}
or 
\begin{equation}
(2\pi)^2P_\mathrm{k}^{-2} G^{-1}R^3 (1+q)^2/q^3.
\label{eq:large}
\end{equation}
\citet{ferrarese00} demonstrated a tight correlation 
between the black hole masses and the velocity dispersions of their host bulges, 
and \citet{merritt01} then derived the relationship 
between central stellar velocity dispersions and black hole masses. 
Based on this relationship, an estimated mass of $(1.9^{+1.0}_{-0.8})\times 10^9$ $M_{\sun}$ 
\citep{noel-storr07} is adopted as the total mass of the two black holes [$M(1+q)$] for 3C 66B, 
where $M_{\sun}$ is the mass of the Sun. 
From equations (\ref{eq:small}) and (\ref{eq:large}), 
the following basic BBH physical parameters were estimated;  
the larger black hole mass of $(1.2^{+0.5}_{-0.2}) \times 10^9$ $M_{\sun}$, 
the smaller black hole mass of $(7.0^{+4.7}_{-6.4}) \times 10^8$ $M_{\sun}$, 
and the separation ($d$) between the black holes of $(1.9\pm0.3) \times 10^{16}$ cm 
[$(6.1^{+1.0}_{-0.9}) \times 10^{-3}$ pc] 
that corresponds to about $53^{+1}_{-10}$ times 
the Schwarzschild radius ($R_s=2GM/c^2$) of $(3.5^{+1.5}_{-0.5}) \times 10^{14}$ cm 
[$(1.1^{+0.5}_{-0.1}) \times 10^{-4}$ pc, $23^{+10}_{-3}$ AU]. 
This estimated separation is much shorter than that of a celestial object recently 
identified as being a potential sub-parsec BBHs by \citet{boroson09}, 
revealing the presence of a very close BBH in the center of \object{3C 66B}.

The decay time of a BBH in the center of \object{3C 66B} due to the gravitational radiation 
is estimated to be $(5.1^{+60.5}_{-2.5})\times 10^2$ years using the 
following equation \citep{peter64,yu02} 
\begin{equation}
5.8\times10^6 \left(\frac{M}{10^8 M_{\sun}} \right)^{-3} \left(\frac{d}{0.01 \mbox{pc}}\right)^4 \frac{1}{q(1+q)}. 
\end{equation}
The estimated decay time is much shorter than the Hubble time. 
The model of galaxy formation based on $\Lambda$CDM indicates 
that there were a large number of galaxy mergers, 
which led to giant elliptical galaxies. 
Centaurus A (CenA), which is the nearby radio galaxy and bright elliptical galaxy, has 
the prominent warped dust lane and gas disk in the central region (e.g., \citealt{baade54}), 
suggesting that Cen A experienced a recent merging event 
during a few $10^8$ years
(\citealt{struve10} and references therein). 
3C66B is a giant elliptical galaxy within the cluster Abell 347 \citep{fanti82} 
and has the prominent face-on dust lane \citep{verdoes99}. 
These indicate that 3C 66B differs from quasars with active merging \citep{volonteri09} 
and is one of nearby radio galaxies that experienced the galaxy merging event.
Within a timescale of $10^8$ years, 
two black holes caused after the galaxy merging event would come closer to each other due to the dynamical friction and gas interaction, 
and then merge into one due to the gravitational-wave radiation 
(e.g., \citealp{begelman80}). 
In conclusion, this shows a possibility that a very close BBH 
in the center of \object{3C 66B} exists 
just before its merger into one black hole.
The detection of strong gravitational-wave radiation is expected 
during the final stage of a black hole merger.

In this Letter, the following assumptions are adopted; 
1) the derived apparent speed of the jet is equal to the bulk flow speed; 
2) the jet axis is parallel to the binary angular momentum and 
is perpendicular to the accretion disk around a black hole 
by which the observed jet is formed; 
3) the BBH has a binary orbit with an orbital eccentricity of zero (this is a circular orbit);
and 4) the spectral index is zero at around our observing frequencies. 
It is theoretically suggested that the pattern speed 
is slower than the bulk flow speed of the jet \citep{lind85}, 
while 
the study on the relativistic bulk motion in about 100 AGNs 
demonstrates that there is no clear distinction 
between these speeds \citep{ghisellini93}. 
For instance, the ratio of these speeds for GRS1919+105 
was estimated to be 0.8 \citep{bodo95}. 
As our estimated proper motion has an error of 15 \% 
that is comparable to this uncertainty, the assumption 1) can be considered reliable 
in this Letter. 
The complexity of mergers and active galactic nuclei (AGNs) feeding may not support the assumptions 2) and 3). 
To evaluate these assumptions, it is necessary to 
monitor the flux variation with higher precision as closely as possible.  
If these assumptions are not entirely inappropriate, however, 
our derived physical parameters are not significantly changed.
Since the synchrotron radiation from the AGN core is expected to be optically thin or peak 
at around our observing frequencies, 
we adopted the assumption 4) in this Letter. 
When $\alpha=0.5$ as compared with $\alpha=0.0$, 
the derived physical parameters of a BBH in \object{3C 66B} are changed at a factor of 1 to 3.
Using the Atacama Large Millimeter/submillimeter Array (ALMA) 
including the Atacama Compact Array (ACA) 
that is used to obtain precise images of celestial objects 
with high angular resolution \citep{iguchi09}, 
the multi-wavelength profile of the periodic flux variation and periodic outburst activities 
of candidate objects can be monitored 
with precision of one mJy or better. 
The spike profile of the periodic flux variation will provide us with the orbital eccentricity of a BBH in \object{3C 66B} 
and the relationship between the jet axis and the binary angular momentum. 
In addition, the multi-wavelength profile will improve 
our derived basic physical parameters of the BBH with 
high precision; 
the orbital radius, the orbital separation and the masses of the two black holes, 
and the decay time.



\acknowledgments

We thank 
M. Saito for his useful comments, 
T. Hasegawa for his comments on data analysis, 
H. Suda for his support with regard to the NMA observations and data reduction, 
and R. Neri for the technical help he provided for the PdBI data reduction. 
This research was partially supported 
by the Ministry of Education, Culture, Sports, Science and Technology (MEXT) of Japan, 
Grant-in-Aid for Young Scientists (B) 17740114, 2005-2007, 
and Grant-in-Aid for Scientific Research (B) 21340044, 2009-2011.



{\it Facilities:} \facility{NMA}, \facility{IRAM (PdBI)}, \facility{VLBA}.

\begin{figure}
\epsscale{.40}
\plotone{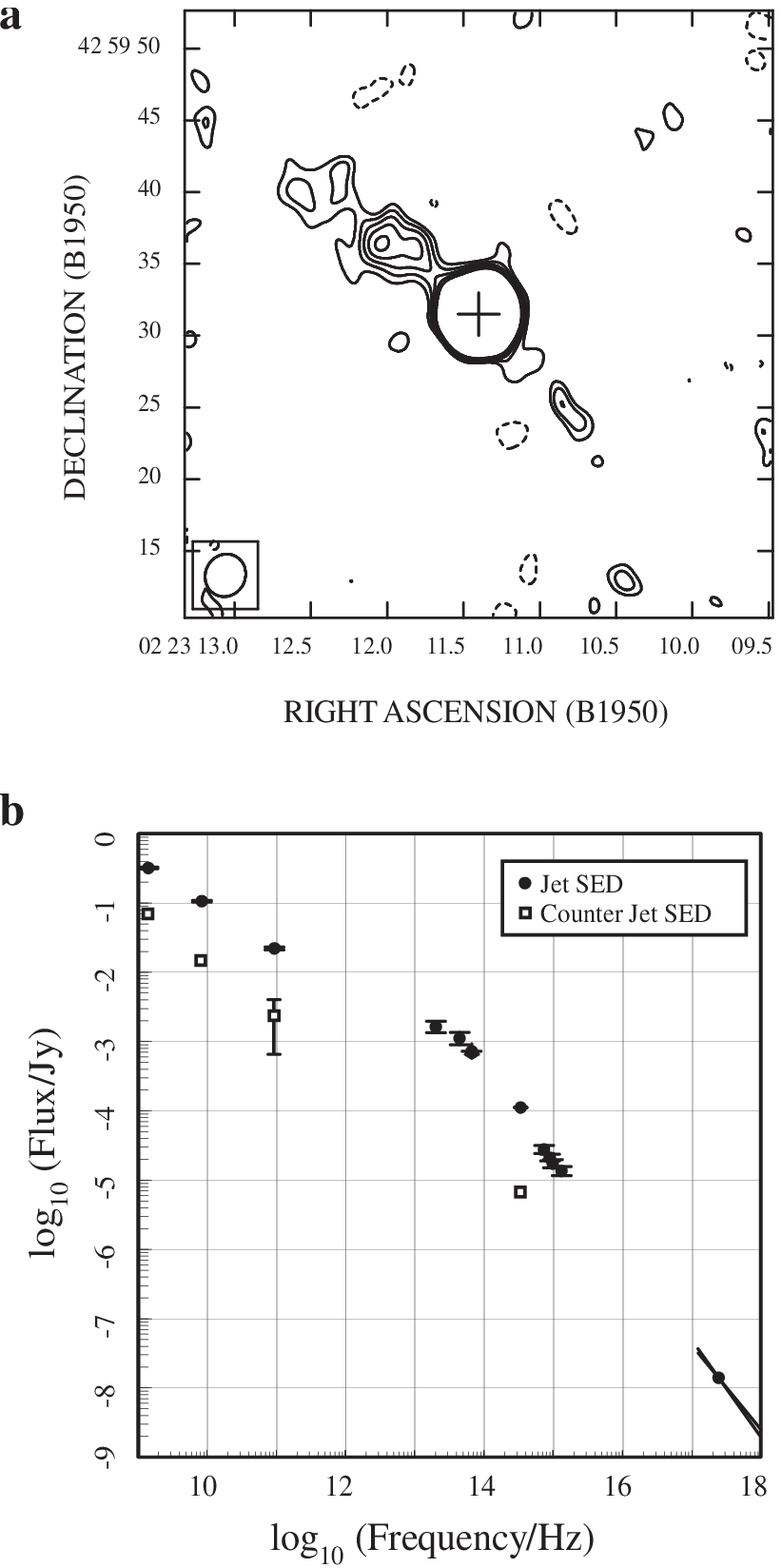}
\caption{
(a) The NMA map of 3C 66B at 93 GHz with the jet and counter jet
with the synthesized beam of 3\farcs0$\times$2\farcs8 with a P.A. -40 degrees, 
with contours at 1.1$\times$(-3, 3, 4, 5, 6, 7) mJy per beam, 
and with a peak flux of 126 mJy per beam. 
(b) Spectral Energy Distributions (SEDs) of the jet (filled circle) 
and the counter jet (open square) in the frequency range of radio to X-ray in 3C 66B. 
The flux densities at several wavelengths 
for the region of the jet and counter jet out to 8 arcsec from the core 
have been tabulated so far \citep{hardcastle01}. 
We estimated an integrated jet flux density of 21$\pm$1 mJy 
and an integrated counter-jet flux density of 2.5$\pm$1.8 mJy at 93.716 GHz, 
and additionally plotted these new points in the spectrums of 
the jet and that of the counter jet, respectively. 
\label{fig1}}
\end{figure}

\begin{figure}
\epsscale{.80}
\plotone{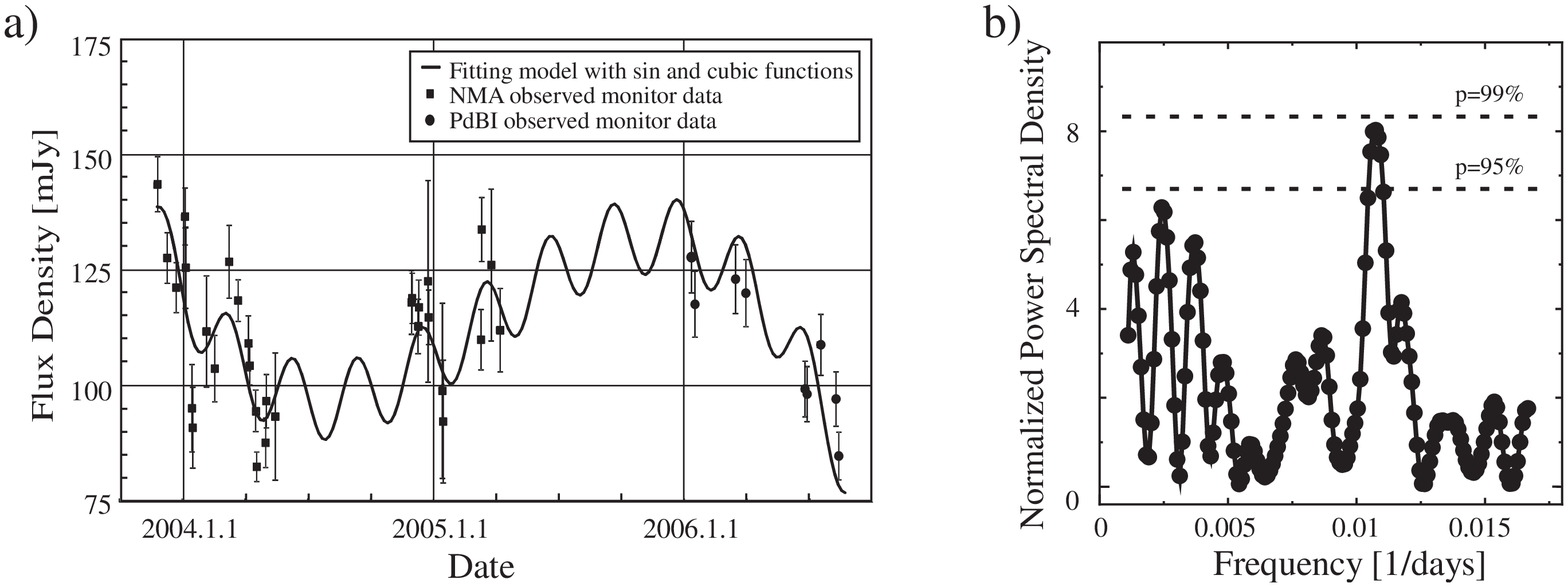}
\caption{
(a) Peak flux monitor data for the millimeter-wavelength core of 3C 66B 
with NMA (filled squares; 93.716 GHz) and PdBI (filled circles; 86.2 GHz). 
(b) Lomb-Scargle periodogram from these flux monitor data. 
We assume that the spectral index of the unresolved core is zero between these two observing frequencies. 
The dotted horizontal lines present the false detection limit 
with the probability of 99 \% and 95 \%, respectively, 
on the presumption that the noise is white noise, 
showing the claimed periodicity of $93\pm1$ days with a 98 \% probability. 
Shown in (a), the claimed periodic flux variation 
with a maximum-to-minimum amplitude ratio of $1.19\pm0.03$ was derived 
by using the minimum Chi-square model fitting 
between the observed data and the sin and cubic functions.
\label{fig2}}
\end{figure}


\begin{figure}
\epsscale{.30}
\plotone{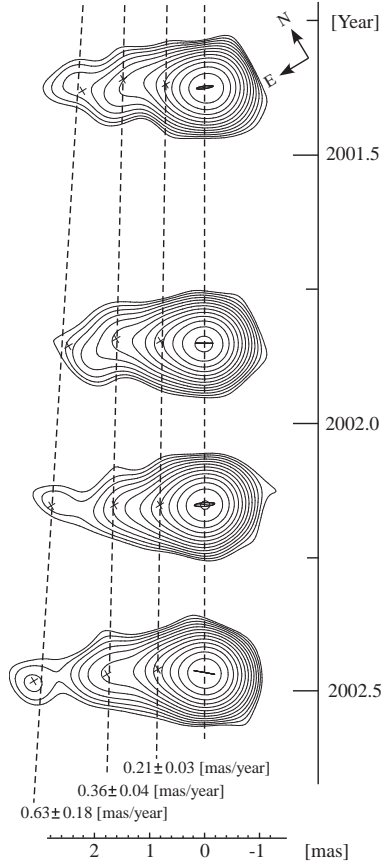}
\caption{
Four 8-GHz VLBI maps of the jet of 3C 66B observed 
using the Very Long Baseline Array (VLBA) 
of the National Radio Astronomical Observatory (NRAO) 
between 2001 March 13 and 2002 June 14 \citep{sudou03}.  
These maps were obtained using a restored synthesized
beam of 0.8$\times$0.8 mas with contours at 
3$\sigma \times (\sqrt{2}, 2, 2\sqrt{2}, ...)$ 
where $\sigma$ is 0.5 mJy per beam, 
and were fitted with one Gaussian component in the core and three point components in the jet.
The proper motions of the three jet components (dashed lines) were determined to be 
0.21$\pm$0.03, 0.36$\pm$0.04 and 0.63$\pm$0.18 mas per year by using the five maps 
(including the map obtained on 2002 February 21; this map has been removed in this figure, however, 
because the date is close to that of the map on 2002 February 8). 
Considering a relativistic emission region directly affected by a BBH, 
we, in this Letter, use the proper motion of the most inner jet, giving 
an apparent speed ($\beta_\mathrm{app}$) of 0.30$\pm$0.04. 
\label{fig3}}
\end{figure}

\begin{figure}
\epsscale{.90}
\plotone{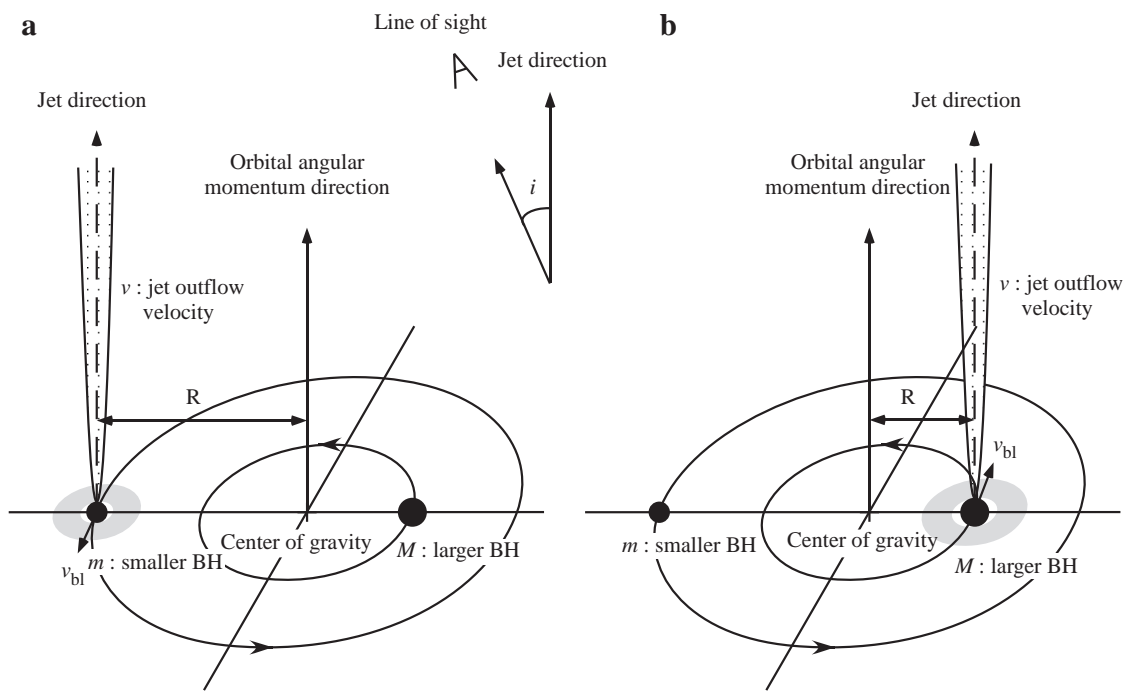}
\caption{
Two schematic geometries of a BBH in the elliptical galaxy 3C 66B. 
We assume that the BBH in 3C 66B has a circular orbit, 
that the jet is linked to the accretion disk around one of the two black holes, 
that the non-thermal radiation propagates outwards 
from the core along the jet with a velocity of $v$, 
and that the jet axis is parallel to the total angular momentum of the binary. 
The observed jet is formed by either (a) the smaller massive black hole 
(with a mass of $m$) 
or by (b) the larger massive black hole (with a mass of $M$).
\label{fig4}}
\end{figure}




\end{document}